# Electric Field Propagation Through Singular Value Decomposition

David Yevick
Department of Physics
University of Waterloo
Waterloo, ON N2L 3G7

**Abstract:** We demonstrate that the singular value decomposition algorithm in conjunction with the fast Fourier transform or finite difference procedures provides a straightforward and accurate method for rapidly propagating electric fields in the one-way Helmholtz formalism.

**Introduction:** One-way electric field propagation can be rapidly implemented in the Fresnel approximation with a variety of numerical methods. However, beams that exhibit a broad angular spectrum or that contain substantial evanescent field components are instead described by the wide-angle [1] or Helmholtz equations. [2] [3] [4] Additionally, if polarization effects are significant, both transverse electric field components must be simultaneously propagated, [5] introducing intrinsic divergences that can only be resolved by specialized procedures such as complex wide-angle Pade approximants. [6] [7]

While various one-way approximations have been suggested for the Helmholtz operator that retain the square-root form of the propagator, [2] a more systematic procedure is afforded by the Lanczos algorithm, which generates an orthogonal basis set for the propagation operator at axial distance $z$ from the set of functions $H^n E(x,y,z), n = 1,2,...,N_{\text{cutoff}}$. [8] [9] The Helmholtz propagation operator is then approximated by its projection onto this restricted basis set. Unfortunately, the procedure converges slowly with the number of basis functions for beams with large angular spreads. [10]

In this paper, an alternative, more general procedure for propagating electric fields in multiple dimensions based on singular value decomposition (SVD) is presented. This analysis is motivated by tensor network algorithms, [11] [12] [13] [14] [15] which, for example, employ SVD to map statistical quantities such as the partition function from a periodic lattice onto a smaller lattice with the identical geometry. Although sophisticated algorithms have been developed to maximize accuracy, [12] [16] [17] [18] a straightforward version of the method that incorporates finite boundary conditions [19] provides sufficient background for most practical applications. Similarly, Helmholtz one-way electric field propagation are here realized by combining the central element of tensor network methods, namely the SVD, with the fast Fourier transform (FFT) and finite difference (FD) methods. The resulting procedure can be immediately adapted to more complex linear wave equations such as the vector wave equation, [5] [6] although in the case that the wave equation contains products of position and derivative operators the FFT procedure entails a real space convolution [20] while the FD method must be precisely formulated [7].

**Computational Method:** Neglecting polarization effects, a monochromatic electric field is described by the scalar wave equation, which can be written in terms of $E(x,y,z) = \mathrm{E}(x,y,z)e^{-i\omega t}$, with $\mathrm{E}(x,y,z)e^{i\omega t}$ the physical electric field amplitude and the vacuum wavenumber $k_0$ as

$$\frac{\partial^2 E}{\partial z^2} = -\frac{\partial^2 E}{\partial x^2} - \frac{\partial^2 E}{\partial y^2} - k_0^2 n^2(x,y,z)E \equiv \left(-\nabla_\perp^2 - k_0^2 n^2(x,y,z)\right)E \equiv -HE \quad (1)$$

The one-way equation, which describes propagation in the forward, $z$, direction namely,

$$\frac{\partial E}{\partial z} = \left(-\nabla_\perp^2 - k_0^2 n^2(x,y,z)\right)^{\frac{1}{2}} E = i\sqrt{H} E \quad (2)$$

possesses the formal solution

$$E(x,y,z+\Delta z) = e^{i\Delta z \left(\nabla_\perp^2 + k_0^2 n^2(x,y,z)\right)^{\frac{1}{2}}} E(x,y,z) = e^{i\Delta z \sqrt{H}} E \quad (3)$$

where $\sqrt{H}$ possesses an imaginary component for evanescent waves. [7]

To implement Eq.(3) numerically, the modified electric field $E$ is first discretized on an uniformly spaced grid of $N \times M$ points (in this paper for simplicity $N = M$). The operator $\nabla_\perp^2 + k_0^2 n^2$ is then represented here in two ways. The first method considers for each grid point in turn, $P \equiv (x_p, y_q)$, a field that is unity at $P$ and zero elsewhere. [20] [21] The FFT is applied to this excitation and the transformed field multiplied by the operator $\nabla_\perp^2$, which is diagonal in wavevector space. After inverse Fourier transforming, the resulting values are entered into a four dimensional tensor $(\mathbf{T})_{mnij} \equiv T(x_m, y_n, x_p, y_q)$ and the diagonal position space operator $k_0^2 n^2(x_p, y_q, \bar{z})\delta_{mp}\delta_{nq}$ is added to $\mathbf{T}$, where $\bar{z}$ denotes an appropriate average value of $z$ over the step length $\Delta z$ (in most cases $\bar{z} = z + \Delta z/2$). These manipulations are clarified by the associated python code for a $z$-invariant refractive index given by, with the formatting conventions of [22] and [23],

```
import numpy as np
n2x = int( numberOfPoints / 2 )
derivativeOperatorX = np.outer( onesValues, np.hstack( [ np.arange( n2x + 1 ), \
-np.arange( n2x - 1, 0, -1 ) ] )**2 )
  derivativeOperatorY = derivativeOperatorX.transpose( )
  propagationMatrix = -( 2 * np.pi / windowWidth )**2 * \
  ( derivativeOperatorX + derivativeOperatorY )
  for xLoop in range( numberOfPoints ):
    for yLoop in range( numberOfPoints ):
      initialField = np.zeros( [ numberOfPoints, numberOfPoints ] )
      initialField[xLoop, yLoop] = 1.0
      fftField = np.fft.fft2( initialField )
```

```
      fftField = propagationMatrix * fftField
      resultField = np.real( np.fft.ifft2( fftField ) )
      for xPrimeLoop in range( numberOfPoints ):
        for yPrimeLoop in range( numberOfPoints ):
          propagationTensor[xPrimeLoop, yPrimeLoop, xLoop, yLoop] = \
          resultField[xPrimeLoop, yPrimeLoop]
          if xPrimeLoop == xLoop and yPrimeLoop == yLoop:
            propagationTensor[xPrimeLoop, yPrimeLoop, xLoop, yLoop] += \
            indexTensor[xPrimeLoop, yPrimeLoop]
```

The second procedure instead employs a 5 point FD stencil for the transverse Laplacian operator and is far more simply programmed:

```
for xLoop in range( numberOfPoints ):
  for yLoop in range( numberOfPoints ):
    propagationTensor[xLoop][yLoop][xLoop][yLoop] = \
    indexTensor[xLoop][yLoop]  - 4 / deltaX**2
    xLoopP1 = np.mod( xLoop + 1, numberOfPoints )
    yLoopP1 = np.mod( yLoop + 1, numberOfPoints )
    propagationTensor[xLoopP1][yLoop][xLoop][yLoop] = 1 / deltaX**2
    propagationTensor[xLoop][yLoopP1][xLoop][yLoop] = 1 / deltaX**2
    propagationTensor[xLoop - 1][yLoop][xLoop][yLoop] = 1 / deltaX**2
    propagationTensor[xLoop][yLoop - 1][xLoop][yLoop] = 1 / deltaX**2
```

The tensor $(\mathbf{T})_{nmpq}$ is then flattened along its first two and last two axes, yielding a $N^2 \times N^2$ matrix $(\mathbf{T})'_{r=(nm),s=(pq)}$ and a SVD routine applied to decompose the matrix into a product **USV**. After truncation, the rows of the (in general complex conjugate) matrix $\mathbf{V}^*$ and the columns of **U** are composed of the first $N_{\text{singular}}$ eigenvectors of $\mathbf{T}'$ while **S** is a diagonal matrix containing the largest $N_{\text{singular}}$ eigenvalues of $\mathbf{T}'$ (in Python however, the elements of **S** are stored in a one-dimensional array).  The accuracy of the procedure is controlled by $N_{\text{singular}}$, and is formally exact for $N_{\text{singular}} = N^2$.

The propagation operator of Eq.(3) is approximated by multiplying either the $q$:th row of **V** or equivalently the $q$:th column of **U** by $\exp(i\Delta z\sqrt{S}_q)$ (often both **V** and **U** are instead multiplied by the square root of the exponential operator).  Simultaneously restoring the rows of $\exp(i\Delta z\sqrt{S})\mathbf{V}$ and the columns of **U** to their original square matrix form yields the two three dimensional tensors **rightPropagationTensor** and **leftPropagationTensor**:

```
from itertools import product
myIterator = np.arange( numberOfPoints )
myCutoffIterator = np.arange( numberOfSingularValues )
flattenedPropagator = np.zeros( [ numberOfPoints**2, numberOfPoints**2 ] )
for index1, index2, index3, index4 in \
product( myIterator, myIterator, myIterator, myIterator ):
  flattenedPropagator[index1 + numberOfPoints * index2, \
  index3 + numberOfPoints * index4] = propagationTensor[index1, index2, index3, index4]
leftPropagationTensor = np.zeros( [ numberOfPoints, numberOfPoints, \
```

```
numberOfSingularValues ] )
rightPropagationTensor = \
np.zeros( [ numberOfPoints, numberOfPoints, numberOfSingularValues ], dtype = complex )

U, S, V = svd( flattenedPropagator )

for index1, index2, index3 in product( myIterator, myIterator, myCutoffIterator ):
  leftPropagationTensor[index1, index2, index3] = \
  U[index1 + numberOfPoints * index2, index3]
  rightPropagationTensor[index1, index2, index3] = \
  V[index3, index1 + numberOfPoints * index2] * \
  np.exp( np.sqrt( S[index3] ) * 1.0j * propagationDistance )
```

The field $E(x,y,z)$ is propagated by first contracting with the first two indices of **rightPropagationTensor** for each value of the third index, thereby expanding the field in a basis of the lowest order $N_{\text{singular}}$ eigenvectors of $H$. This yields an array of $N_{\text{singular}}$ elements corresponding to the amplitudes of the projections of the propagating field at position $z$ onto the lowest eigenvectors. This coefficient array is finally contracted with the third index of **rightPropagationTensor** and the resulting matrices summed to obtain $E(x,y,z+\Delta z)$.

```
myIntermediateResult = np.zeros( numberOfSingularValues, dtype = complex )
for index1, index2, index3 in product( myIterator, myIterator, myCutoffIterator ):
  myIntermediateResult[index3] += rightPropagationTensor[index1, index2, index3] * \
  field[index1, index2]
field = np.zeros( [numberOfPoints, numberOfPoints], dtype = complex )
for index1, index2, index3 in product( myIterator, myIterator, myCutoffIterator ):
  field[index1, index2] += leftPropagationTensor[index1, index2, index3] * \
  myIntermediateResult[index3]
```

If necessary, the electric field can be multiplied by an absorber function after each propagation step or set of steps to reduce the component of the propagating electric field close to the computational window edges. [24]

If the refractive index does not vary with $z$, then since typically $N_{\text{singular}} \ll N^2$, once the tensors **leftPropagationTensor** and **rightPropagationTensor** have been constructed in the first step of the procedure, subsequent propagation steps only involve contraction with these two tensors and therefore require minimal computational resources. Small refractive index variations can often be simply incorporated perturbatively by multiplying the field by $\exp(i\Delta z k_0 \delta n)$ where $\delta n$ represents the difference between the refractive index averaged over the current propagation step and the average at the initial propagation step. In contrast, significant axial refractive index changes necessitate repeated resource intensive SVD operations.

While the analysis of this paper is framed in the context of the scalar wave equation, the method is general and can be equally applied to, e.g. vector wave equations at any level of approximation. In this case, $H$ typically contains products of refractive index and derivative operators that are typically discretized by the FD method or FFT:s together with a convolution.

**Results:** To illustrate the SVD procedures, a $z$-travelling Gaussian beam with a $4.0 \mu m$ width, wavelength $\lambda_0 = 1.3 \mu m$ and an initial offset of $5.0 \mu m$ in the vertical direction is propagated through a $z$-invariant quadratic refractive index profile given by

$$n^2(r) = 1.45 \left(1 - 0.1 \left(\frac{\min(r, 25)}{25}\right)^2\right) \quad (4)$$

A $40 \times 40$ point grid with a $1.0 \mu m$ point separation was employed in the calculation. In this profile, classical paraxial rays describe sinusoidal trajectories with a $496.7 \mu m$ period. Accordingly, the electric field amplitude contours obtained after 496 steps of length $1.0 \mu m$ with the standard FFT based Fresnel propagation method [25] and a reference refractive index of 1.442 are displayed as the solid lines of Fig. 1. Superimposed on these curves are the corresponding results of two Helmholtz SVD calculations with $N_{\text{cutoff}} = 80$ of which the first employs the FD (dotted line) and the second the FFT approach (dashed line) of the previous section. As expected, the center of the beam is displaced during propagation from $+5 \mu m$ to $-5 \mu m$ before returning to its initial position. The results of the three methods in Fig. 1 are nearly identical apart from slight divergences associated with the differing propagation operators and the elevated FD discretization error. While this result verifies the feasibility of the SVD technique, its efficiency is governed by $N_{\text{cutoff}}$ and hence the desired level of accuracy. The latter is, in turn, intrinsically problem-dependent, precluding a straightforward comparison with other procedures.

**Discussion and Conclusions:** We have demonstrated that SVD decomposition can be employed to propagate electric fields both simply and accurately. The procedure enables the direct evaluation of the one-way Helmholtz propagation operator and is therefore applicable to fields with both large angular spreads and evanescent components. The tradeoff between accuracy and computational efficiency is controlled through a single parameter, the number of basis functions, such that the discretized representation of the square-root operator becomes formally exact when this parameter equals the number of computational grid points. Further, vector field propagation can be analyzed either in the FD approximation or by performing appropriate convolutions to form the matrix representations of products of derivative and position operators.

On the other hand, depending on the desired degree of accuracy, the SVD procedure can require far greater computational resources than standard FFT or FD propagation algorithms, although iterative SVD methods exist that can potentially reduce the computational overhead. [26] Furthermore, while refractive index distributions that are nearly invariant along the optical axis can be accurately incorporated by repeated multiplications by additional position space operators, for pronounced longitudinal variations the full set of basis functions must be frequently recomputed, severely degrading computational efficiency.

**Acknowledgements:** The Natural Sciences and Engineering Research Council of Canada (NSERC) is acknowledged for financial support.

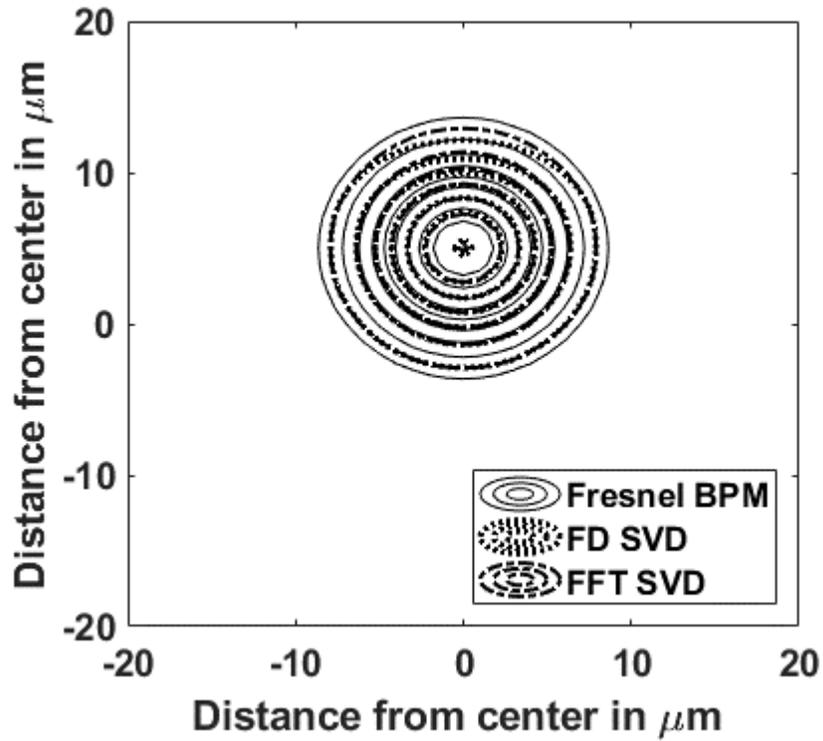

*Figure 1: The electric field amplitude after a 5µm wide Gaussian beam is propagated one ray period through a parabolic fiber refractive index profile with the Fresnel FFT beam propagation method (sold lines), the Helmholtz FD SVD approach (dotted line) and the Helmholtz FFT based SVD technique (dashed line).*